\documentclass[12pt]{article}
\usepackage{epsf,amsfonts}

\epsfverbosetrue
\makeatletter
\def\@citex[#1]#2{\if@filesw\immediate\write\@auxout
        {\string\citation{#2}}\fi
\def\@citea{}\@cite{\@for\@citeb:=#2\do
        {\@citea\def\@citea{,}\@ifundefined
        {b@\@citeb}{{\bf ?}\@warning
        {Citation `\@citeb' on page \thepage \space undefined}}
        {\csname b@\@citeb\endcsname}}}{#1}}
\newif\if@cghi
\def\cite{\@cghitrue\@ifnextchar [{\@tempswatrue
        \@citex}{\@tempswafalse\@citex[]}}
\def\citelow{\@cghifalse\@ifnextchar [{\@tempswatrue
        \@citex}{\@tempswafalse\@citex[]}}
\def\@cite#1#2{{\if@cghi\unskip$\null^{#1}$\else #1\fi\if@tempswa\typeout
        {warning: optional citation argument ignored: `#2'} \fi}}

\def\@biblabel#1{$\null^{#1}$}

\makeatother

\usepackage{fontenc,indentfirst, delarray,amsfonts,amsmath,amssymb}
\usepackage{graphicx}
\DeclareGraphicsExtensions{ps,eps,ps.gz}

\begin{document}
\font\twelve=cmbx10 at 13pt
\font\eightrm=cmr8
\baselineskip 18pt

\def\ccc{{\double C}}
\def\aa{{\cal A}}
\def\dd{{\cal D}}
\def\hh{{\cal H}}
\def\jj{{\cal J}}
\def\oo{{\cal O}}
\def\lb{\left[}
\def\rb{\right]}
\def\lp{\left(}
\def\rp{\right)}
\def\bbb{\begin{equation}}
\def\eee{\end{equation}}
\def\pp{\pmatrix}

\hsize 17truecm
\vsize 24truecm

\font\thirteen=cmbx10 at 13pt
\font\ten=cmbx10
\font\eight=cmr8

\baselineskip 20pt

\renewcommand{\baselinestretch}{1.6}

\thispagestyle{empty}

\centerline{\ten Centre de Physique
Th\'eorique\footnote{\eight Unit\'e Propre de Recherche
7061}, CNRS Luminy, Case 907}

\centerline{\ten F-13288 Marseille -- Cedex 9}

\vskip 2truecm

\centerline{\thirteen DISTANCES IN FINITE SPACES }
\centerline{\thirteen FROM NONCOMMUTATIVE GEOMETRY.}
\bigskip
\bigskip

\centerline{ 
{\bf 
Bruno IOCHUM
\footnote{\eight and Universit\'e de Provence, iochum@cpt.univ-mrs.fr},
Thomas KRAJEWSKI
\footnote{\eight and Universit\'e de Provence, tkrajews@cpt.univ-mrs.fr},
Pierre MARTINETTI
\footnote{\eight and Universit\'e de Provence, martinet@cpt.univ-mrs.fr},
 }}
\vskip 2truecm
\centerline{\bf Abstract} 
\medskip
Following the general principles of noncommutative geometry, it is possible to 
define a metric on the space of pure states of the noncommutative algebra generated by 
the coordinates.  This metric generalizes the usual Riemannian one. We 
investigate some general properties of this metric in the finite commutative 
case which corresponds to a metric on a finite set and also give some examples 
of computations in both commutative and noncommutative cases. 

\vskip 2truecm

PACS-99: 04.60.Nc Lattice and discrete methods. \\
\bigskip

\noindent Number of figures: 2
\bigskip

\noindent December 1999

\noindent CPT-99/P.3927
\bigskip

\noindent anonymous ftp : ftp.cpt.univ-mrs.fr

\noindent web : www.cpt.univ-mrs.fr

\newpage

\section{Introduction.}

Though particle experiments are going further in energy and consequently deeper in the structure of matter, the geometric structure of space time is
still unknown. Classical differential geometry does not allow to take seriously into account both general relativity  and quantum mechanics since the
latest renounces intuitive geometric concepts while the first grounds its decription of gravitation on purely geometric concepts. Noncommutative
differential geometry \cite{connes} gives a mathematical framework for a geometric understanding of fundamental interactions. Saying geometric
understanding, one would like to say clearer understanding: for instance the noncommutative standard model \cite{gravity,bridge,iochum} gives
a geometric interpretation of the Higgs field together with an estimation on the mass of the corresponding boson. 

But it is still difficult to draw an intuitive picture of a noncommutative space, as one can do for an euclidean 
or even riemannian space. A noncommutative space is described by a $C^*$-algebra $\mathcal{A}$, a faithful representation of $\mathcal{A}$ over a
Hilbert space $\mathcal{H}$, and an operator $D$ acting on $\mathcal{H}$.  $D$ has a compact resolvant and is possibly unbounded. To be precise, the 
algebra is restricted to the norm closure of the set of elements $a\in \mathcal{A}$ such that $[D,\pi(a)]$ is bounded.
 
A distance is then defined on $\mathcal{S}(\mathcal{A})$, state space of $\mathcal{A}$, by
\begin{equation}
\label{intro}
d({\Phi} , {\Psi}) = \underset {a \in \mathcal{A}} {\sup}  \, \{  \, 
|{\Phi}(a) - {\Psi}(a)|  \;\,  / \;\, \|[D,a]\| \leq 1 \},
\; \forall \, {\Phi},{\Psi} \in \mathcal{S}(\mathcal{A}).
\end{equation}

If $\mathcal{A}$ is commutative, pures states correspond to characters. Thanks to Gelfand construction, they are interpreted as points, and  $\mathcal{A}$ is the algebra of functions over these
points. When $\mathcal{A}$ is not commutative, this interpetation is no more possible but the distance formula between pure states remains unchanged. 

There is not always a clear physical interpretation of this distance. When $\mathcal{A}
$ is the algebra of smooth functions over a riemannian spin  manifold, $\mathcal{H}$ the space of $L^2$-spinors and $D$ the classical Dirac
operator, then the noncommutative distance coincides with the geodesic distance. When $\mathcal{A}$ is tensorised by an internal algebra, for instance 
the diagonal 2x2 matrices, then one obtains a space of two sheets with geodesic distance over each of them. Noncommutative distances have been considered 
\cite{bimonte,dimakis1,dimakis3} in the case of finite algebras. Moreover, the classical distance in one dimensional lattices can be obtained via the noncommutative approach \cite{atzmon,dimakis2}.

In  Refs.\ \citelow{rieffel1,rieffel2}, the problem is introduced in a general framework: let $L$ be a Lipschitz seminorm on a partially ordered real vector
space $A$. $L$ determines a metric $\rho_L$ in the state space $\mathcal{S}(A)$:
$$
\rho_L(\mu , \nu) = \underset {a \in \mathcal{\mathcal{A}}} {\sup}  \, \{  \, 
|\mu(a) - \nu(a)|  \;\,  / \;\, L(a) \leq 1 \}\quad \text { with } \mu,\nu \in \mathcal{S}(A).
$$
$\rho_L$ determines a Lipschitz seminorm $L_{\rho_L}$ over the space $Af[\mathcal{S}(A)]$ of affine
functions on $\mathcal{S}(A)$:
$$
L_{\rho_{L}}(f) =  \underset {{\mu,\nu} \in \mathcal{S}(\mathcal{A})} {\sup} \{ {|{f(\mu) - f(\nu)}| \over {\rho(\mu, \nu)}} / \mu \neq \nu\}\quad \text{with }
f\in Af[\mathcal{S}(A)]. 
$$
As $A$ is isomorphic to a dense subspace of $Af[\mathcal{S}(A)]$, the
question is: under which conditions one has $L_{\rho_L}=L$ ? Answers
are given in Ref.\ \citelow{rieffel2}. In the noncommutative framework,
taking $L(a)=\|[D,a]\|$, the question becomes: how to characterize the
metrics $\rho$ coming from a Dirac operator ?

In this paper, we carry out the calculation  of distances in spaces associated to finite dimensional algebras. It seems natural to restrict to finite dimensional
representation of these algebras since, in the noncommutative approach of the standard model, the internal space is the space of fermions (more mathematical
arguments to avoid infinite representations of finite dimensional algebras can be found in Ref.\ \citelow{krajew2}).  Therefore, $\mathcal{A}$ is a direct 
sum of
$k$ matrix algebras, since any involutive algebra over $\mathbb{C}$ which admits a faithful finite dimensional representation in a Hilbert space is a direct sum of
matrix algebras. For $k=1$, the simpliest interesting case is $\mathcal{A}=M_2(\mathbb{C})$. The associated space is 
a fiber space whose base has only one point and the fiber is $\mathbb{C}^2.$ For $k=2$, we study the noncommutative space associated to 
$\mathcal{A}=M_p(\mathbb{C})\oplus\mathbb{C},\, p\in \mathbb{N}$. This is a two point space with fiber $\mathbb{C}^p$ over one of the point. Some applications can be found in Ref.\ \citelow{rovelli}, where $\mathcal{A}=M_2(\mathbb{C})\oplus \mathbb{C}$ 
is used to build a first model of quantized spacetime.

For $k\geq 3$, we restrict to commutative algebras. Then $\mathcal{A}=\mathbb{C}^k$ and ${S}(\mathcal{A})$ 
is simply a set of $k$ points. We choose $\mathcal{H}=\mathbb{C}^k$. For the three point space with any
real selfadjoint operator $D$ and the $n$ point space with some particular operators $D$, we explicitly compute distances. To find a Dirac which
gives a desired metric, it is enough to inverse formula. This is not possible in the four point case for we show that generic distances are roots of
polynomials which cannot be solved by radicals. 
A possible solution consists in modifying our definition of commutative spaces. Using a slightly more complicated representation of 
$\mathcal{A}$ over a space $\mathcal{H}'$ larger than $\mathcal{H}$, one shows that there always exists an operator $D'$ giving the desired distances.
Moreover $(\mathcal{A},\mathcal{H}',D')$ is a real spectral triple which fulfills all the axioms of noncommutative geometry. 
 
\section{General result and notations.}

All along this paper, $\mathcal{A}$ is a unital $C^*$-algebra represented in a Hilbert space $\mathcal{H}$. 
$D$ is a selfadjoint operator on $\mathcal{H}$ which does not belong to the commutant of $\mathcal{A}$. Its components are 
$D_{ij}$, $1\leq i,j \leq n$. $\mathcal{S}(\mathcal{A})$ is the set of pure states of $\mathcal{A}$. 
The distance associated to the triplet $(\mathcal{A},\mathcal{H},D)$ by (\ref{intro}) is denoted by $d$. 
 $\\$

{\bf Lemma 1.}
{\it  Let $\mathcal{A}_{+}$ be the subset of positive elements of $\mathcal{A}$, and $\Phi,\Psi \in \mathcal{S}(\mathcal{A})$. 
\begin{equation*}
d({\Phi} , {\Psi}) = \underset {a \in \mathcal{A}_{+}} {\sup}  \, \{  \, 
|{\Phi}(a) - {\Psi}(a)|  \;\,  / \;\, \|[D,a]\| = 1 \}.
\end{equation*}}

{\it Proof.}
Let $\theta\doteq \text{arg}\left(({\Phi}-{\Psi})(a_0)\right)$ where $a_0\in \mathcal{A}$ reaches the supremum in (\ref{intro}), namely:
\begin{equation*}
\|[D,a_0]\| \leq 1{\text { and }}\;  |({\Phi} - {\Psi})(a_0)| = 
\text{dist}({\Phi},{\Psi}).
\end{equation*}  

The supremum is also reached for the selfadjoint element  $b_0={1\over 2}({a_0e^{-i\theta} + a_0^*e^{i\theta}})\in \mathcal{A}$ since
\begin{eqnarray*}
&\|[D,b_0]\| \leq  {{\|[D,a_0]\|}\over 2} + {\|[D,a_0^*]\| \over 2} \leq 1,&\\
& |({\Phi} - {\Psi})(b_0)|=|{\text{dist}({\Phi},{\Psi}) \over 2}  + {{\overline{\text{dist}
( {{\Phi} , {\Psi}})}}\over 2}|= \text{dist}({\Phi},{\Psi}).&  
\end{eqnarray*}
The same is true for $c_0= b_0 + \|b_0\|I \in \mathcal{A_+}$, so we restrict to $\mathcal{A}_+$.   

Suppose now $\|[D,c_0]\| <1$. Take $e_0\doteq {c_0 \over \|[D,c_0]\|}\in \mathcal{A}_+$, then 
$$\|[D,e_0]\|=1 \quad \text{ and } \quad |{\Phi}(e_0) - {\Psi}(e_0)| = {|{\Phi}(c_0) - {\Psi}(c_0)| \over \|[D,c_0]\|} > |{\Phi}(c_0) - {\Psi}(c_0)|,$$ 
which is impossible since $c_0$ is chosen to reach the supremum. So $\|[D,c_0]\|=1$.

If  the supremum is not reached, the proof uses a sequence $\{a_n\}$ of elements of $\mathcal{A}. \blacksquare$ 
$\\$

{ \bf $\bullet$ Once for all, any element $a\in\mathcal{A}$ that appears in a proof is selfadjoint.}

$\bullet$ The canonical basis of $\mathbb{C}^n$ ( or $\mathbb{R}^n$ in case) is denoted by $|1\rangle, |2\rangle,...,|n\rangle$.

$\bullet$ When $\mathcal{A}=M_n(\mathbb{C})$, a pure state $\omega_{\xi}$ is determined by a normalized vector $\xi \in \mathbb{C}^{n}:\; \omega_{\xi}(a) = \xi^* a \, \xi$, $\forall a\in 
M_n(\mathbb{C}).$
Two normalized vectors determine the same pure state
if and only if they are equal up to a phase. In other terms,
$\mathcal{S}(M_n(\mathbb{C}))=\mathbb{C}P^{n-1}$. 

$\bullet$ For any unitary operator $U$ of $\mathcal{H}$, the jauge transformed of $\omega_{\xi}$ is $\tilde{\omega}_{\xi}(a) 
\doteq \omega_{\xi}(UaU^{-1})$, $\forall a \in \mathcal{A}$. If $\tilde{D}\doteq U^{-1}DU$, then $d_{\tilde{D}}(\tilde{\omega_{\xi}}, \tilde{\omega_{\zeta}}) =
d_{D}(\omega_{\xi},\omega_{\zeta})$ for 
$${\underset{a\in \mathcal{A}}{\sup}}\left\{ |(\tilde{\omega}_{\xi}-\tilde{\omega}_{\zeta})(a)| \,/\,\|U^{-1}DU,a]\|=1 \right\} = 
 {\underset{UaU^{-1}\in \mathcal{A}}{\sup}} \left\{ |(\omega_{\xi}-\omega_{\zeta}) (UaU^{-1})|\,/\,\|[D,UaU^{-1}]\|=1 \right\}.$$ 

\section{One point space.}
The first non trivial example with a single matrix algebra is $\mathcal{A}=M_2(\mathbb{C})$, represented in $\mathcal{H}=\mathbb{C}^2$ by $\mathcal{A} \ni a=\left( \begin{array}{cc} a_{11} & a_{12} \\ a_{21} & a_{22} 
\end{array} \right)$. $D$ is a $2\times 2$ selfadjoint matrix with two real and distinct eigenvalues $D_1,D_2$. $\mathcal{S}(\mathcal{A})= \mathbb{C}P^{1}$
is isomorphic to the sphere $S^2$. 
An explicit one to one correspondence is the Hopf fibration \cite{nakahara}: $\xi =(\xi_1,\xi_2) \in \mathbb{C}P^1$ is associated to 
$(a_{\xi},b_{\xi}, c_{\xi}) \in S^2\,$ by 
$$a_{\xi}\doteq 2 \mathcal{R}e(\xi_1 \bar{\xi}_2), \quad b_{\xi}\doteq 2 \mathcal{I}m(\xi_1 \bar{\xi}_2)\,  \text{ and } \, c_{\xi}\doteq  |\xi_1|^2 -|\xi_2|^2.$$

{\it {\bf Proposition 2.}

\begin{eqnarray*}
 d(\omega_{\xi},\omega_{\zeta}) &=&
 {2\sqrt{ 1-|\langle\xi,\zeta\rangle|^{2}}\over{|D_1-D_2|}} \quad \text{if }\,  c_{\xi}=c_{\zeta},\\
                               &=&  +\infty \quad \text {otherwise.}
 \end{eqnarray*}}

{\it Proof.}
Let $U$ be the unitary operator of $\mathcal{H}$ such that
$\tilde{D}\doteq U^{-1}DU = \text{diag}(D_1,D_2)$. 
A direct computation yields $\|[\tilde{D},a]\|=|a_{12}||D_1-D_2|$, thus the norm condition in (\ref{intro}) becomes
\begin{equation}
\label{onenorme}
|a_{12}|\leq {1 \over |D_1-D_2|}.
\end{equation}
Furthermore, $(\omega_{\xi} -\omega_{\zeta})(a)=
\xi^*a\xi-\zeta^*a\zeta ={\underset{i,j=1}{\overset{2}{\Sigma}}} a_{ij}(\bar{\xi_i}\xi_j-\bar{\zeta_i}\zeta_j).$ 

If $|\xi_1|\neq|\zeta_1|$, 
then $a$ with all coefficients zero, except $a_{11}=L$, verifies the norm condition (\ref{onenorme}) for any $L \in \mathbb{R}^+$, and 
$|(\omega_{\xi} -\omega_{\zeta})(a)|=
 L||\xi_1|^2-|\zeta_1|^2|.$ Thus ${d}_{\tilde{D}}(\omega_{\xi},\omega_{\zeta})=+\infty$.

If $|\xi_1|=|\zeta_1|$, then $|\xi_2|=|\zeta_2|$ since $\|\xi\|=\|\zeta\|=1$. So $c_{\xi}=c_{\zeta}$ and  
\begin{equation}
\label{onedist}
|(\omega_{\xi} -\omega_{\zeta})(a)|=
|2\mathcal{R}\text{e} \left(a_{12}(\bar{\xi_1}\xi_2-\bar{\zeta_1}\zeta_2)\right)|\leq 2|a_{12}||(\bar{\xi_1}\xi_2-\bar{\zeta_1}\zeta_2)|.
\end{equation}
As any vector of $\mathbb{C}P^{n-1}$ is defined up to a phase, we assume that $\xi_1=\zeta_1$ is real. Let $\theta_{\xi}\doteq \text{arg}(\xi_2)$ and
$\theta_{\zeta}\doteq \text{arg}(\zeta_2)$. Take $a_{11}=a_{22}=0$ and $\text{arg}(a_{21})={1 \over 2}(\pi - \theta_{\xi}- \theta_{\zeta}).$
Then 
$$
|\mathcal{R}\text{e} \left(a_{12}(\bar{\xi_1}\xi_2-\bar{\zeta_1}\zeta_2)
\right)|=
|a_{12}|\, y\xi_1\, |\xi_2|\, |2\sin({{\theta_{\xi}-\theta_{\zeta}}\over 2})|
=|a_{12}||\bar{\xi_1}\xi_2-\bar{\zeta_1}\zeta_2|.
$$
$a$ reaches the upper bound in (\ref{onedist}) and verifies the norm condition (\ref{onenorme}) as far as one chooses $|a_{12}|={1\over {|D_1-D_2|}}.$ So 
${d}_{\tilde{D}}(\omega_{\xi},\omega_{\zeta})={2\over {|D_1-D_2|}}|\bar{\xi_1}\xi_2-\bar{\zeta_1}\zeta_2|$.
An intrinsic formulation is found by writing
$$
\xi\xi^* -\zeta\zeta^* =\left(
\begin{array}{cc}
0                                      & \xi_1\bar{\xi_2}-\zeta_1\bar{\zeta_2} \\
\xi_2\bar{\xi_1}-\zeta_2\bar{\zeta_1}  & 0                                     \\
\end{array} \right),
$$
$\text{Tr}(\xi\xi^* -\zeta\zeta^*)^2 = 2|\bar{\xi_1}\xi_2
-\bar{\zeta_1}\zeta_2|^2.$
Developing the trace yields $\,1 - |\langle\xi,\zeta\rangle|^{2} 
=|\bar{\xi_i}\xi_1-\bar{\zeta_i}\zeta_1|^2y$, thus
$${d}_{\tilde{D}}(\omega_{\xi},\omega_{\zeta})={2\over {|D_1-D_2|}}\sqrt{ 1-|\langle\xi,\zeta\rangle|^{2}}.$$ 
Finally, ${d_{\tilde{D}}}(\omega_{\xi},\omega_{\zeta})={d_{\tilde{D}}}(U^{-1}\xi,U^{-1}\zeta)$ which is equal to $d_{D}(\omega_{\xi},\omega_{\zeta})$ as seen before. $\blacksquare$
\medskip
 
We say that two states $\omega_{\xi},\omega_{\zeta}\in\mathcal{S}(\mathcal{A})$ are at the same altitude if $c_{\xi}=c_{\zeta}$. By an easy calculation, 
for two such states, $d(\omega_{\xi},\omega_{\zeta})={2\over {|D_1-D_2|}}\sqrt{(a_{\xi} - 
a_{\zeta})^2+(b_{\xi}-b_{\zeta})^2}$. In other terms, up to a constant factor, $d$ is nothing but the 
euclidean distance restricted to plans of constant altitude. The distance between to planes of different altitudes is infinite. 

In a one point space with a fiber of higher dimension than $\mathbb{C}^2$, one needs an explicit formula 
for the norm of a selfadjoint $n\times n$ complex matrix. These cases are not studied here.

\section{Two point space.}
Consider the algebra $\mathcal{A}=M_{n}(\mathbb{C})\oplus\mathbb{C}$ represented on 
$\mathcal{H}=\mathbb{C}^{n}\oplus\mathbb{C}$ by 
 $\mathcal{A}\ni a=\left(\begin{array}{cc}x & 0\\ 0 & y\end{array}\right)$, $x \in M_n(\mathbb{C}),\, y\in \mathbb{C}$. 
The simpliest interesting operator is $D=\left(\begin{array}{cc}0 & m\\ m^* & 0\end{array}\right)$, 
with $m\in\mathbb{C}^{n}$ a non zero column vector.
$\mathcal{S}(\mathcal{A})$ is the union of the single pure state of
$\mathbb{C},\,  \omega_0\doteq Identity$, with $\mathcal{S}(M_n(\mathbb{C}))$. 
$\\$

{\it {\bf Proposition 3.}
\begin{eqnarray*}
&&{\bf  i)} \left\{
\begin{array}{l}
d(\omega_{\xi},\omega_0)=\frac{1}{\|m\|}
\quad\text{if }\, \xi\text{ and m are colinear,}\\
d(\omega_{\xi},\omega_0)=+\infty\quad\text{otherwise.}
\end{array}
\right.\\   
&&{\bf ii)} \left\{
\begin{array}{l}
d(\omega_{\xi},\omega_{\zeta})=\frac{2}{\|m\|}\sqrt{ 1-1|\langle\xi,\zeta\rangle|^{2}}
\quad\text{ if }\;(\xi-\zeta e^{i\theta}) \text { and m are colinear for some } \theta \in [0,2\pi[,\\
d(\omega_{\xi},\omega_{\zeta})=+\infty\quad\text{otherwise.}
\end{array}
\right. 
\end{eqnarray*} }

{\it Proof.}
We may assume that $\|m\|=1$, for dividing  $D$ by $ \|m\|$ means multiplying distances by $\|m\|$. Thus, there is a unitary operator 
$u\in M_n(\mathbb{C})$ such that  $m=u|1\rangle$. With $ U\doteq \left(
\begin{array}{cc}
u &  0\\
0 &  1 
\end{array}
\right)
$, $\tilde{D}\doteq U^{-1}DU$ and $z\doteq x-yI_{n}\,$ (selfadjoint by lemma 1) one has
\begin{equation}
\label{NGCnorme}
\|[\tilde{D},a]\|^2=\underset{i=1}{\overset{n}\Sigma}|{z}_{i1}|^2,
\end{equation}
since
 $$\|[\tilde{D},a]\|^2
=\| \left( \begin{array}{cc}
0  & -z|1\rangle \\ 
\langle1|z  &   0
\end{array}\right) \|^2
=\|\left(\begin{array}{cc}z|1\rangle\langle1|z^* & 0 \\ 0                   &\langle1|zz^*|1\rangle \end{array} \right)\|,$$
and 
$$\|z|1\rangle\langle1|z^*\| =\| z|1 \rangle\|^2=\underset{i=1}{\overset{n}\Sigma}|\tilde{z}_{i1}|^2\, ,\quad
|\langle1|zz^*|1\rangle|=\underset{i=1}{\overset{n}\Sigma}|\tilde{z}_{i1}|^2.$$
 
Note that $\, \|[\tilde{D},a]\|^2=1$ implies $\; |z_{i1}| \leq 1 \quad \forall i\in \{1,n\}$.

{\it i)} $(\omega_{\xi} -\omega_0)(a)=
\xi^*x\xi-y =
\xi^*z\xi={\underset{i,j=1}{\overset{n}{\Sigma}}}\bar{\xi_i} z_{ij}
\xi_j.$ 

Assume $\xi_{k}\neq 0$ for $k\in\{2,n\}$. Take the matrix $z$ with all coefficients zero except $z_{kk}=L\in\mathbb{R}^+$. By ({\ref{NGCnorme}), $z$ satisfies the norm condition
of formula (\ref{intro}) and $|(\omega_{\xi} -\omega_0)(a)|=|\xi_k|^2L$. Thus $d_{\tilde{D}}(\omega_{\xi},\omega_{\xi_0})=+\infty$.

Assume $\xi_i =0,\,  \forall i \in \{2,n\}$: there is constant $\theta$ such that $\xi = e^{i\theta}|1\rangle$. So $|(\omega_{\xi} -\omega_0)(a)|=|z_{11}|
\leq 1$.  This upper bound is reached by $z$ with all coefficients zero except $z_{11}=1$.

{\it ii)} $(\omega_{\xi} - \omega_{\zeta})(a)={\underset{i,j=1}{\overset{n}{\Sigma}}} z_{ij}(\bar{\xi_i}        \xi_j - \bar{\zeta_i} \zeta_j).$ 

Assume $(\bar{\xi_{p}} \xi_{l} - \bar{\zeta_{p}} \zeta_{l})\neq 0$ for $p,l \in \{2,n\}$. The proof is similar as $i)$ with $z=0$
except $z_{pl}=L$.

Assume $\bar{\xi_{i}} \xi_{j} - \bar{\zeta_{i}} \zeta_{j}=0 \;\, \forall i,j \in \{2,n\}.$ This is equivalent to $\xi_i=\zeta_i e^{i\theta}$, with 
$\theta$ a constant. In other words, $(\xi-\zeta e^{i\theta}) \sim|1\rangle$.  Furthermore, since $\|\xi\|^2=\|\zeta\|^2=1$, one has
$|\xi_1|=|\zeta_1|.$ Thus 
\begin{eqnarray}
\nonumber
&|{\underset{i,j=1}{\overset{n}{\Sigma}}} z_{ij} (\bar{\xi_i}\xi_j- \bar{\zeta_i}\zeta_j)|= 2|{\underset{i=2}  {\overset{n}{\Sigma}}} 
\mathcal{R}\text{e}\left(z_{i1}(\bar{\xi_i}\xi_1-\bar{\zeta_i}\zeta_1)\right)|
\leq 2|{\underset{i=2}  {\overset{n}{\Sigma}}}z_{i1}(\bar{\xi_i}\xi_1-\bar{\zeta_i}\zeta_1)|&\\ 
\label{above}
&\leq 2\sqrt{{\underset{i=2} {\overset{n}{\Sigma}}}|z_{i1}|^2} 
       \sqrt{{\underset{i=2} {\overset{n}{\Sigma}}}|\bar{\xi_i}\xi_1-\bar{\zeta_i}\zeta_1|^2}
\leq 2 \sqrt{{\underset{i=2} {\overset{n}{\Sigma}}}|\bar{\xi_i}\xi_1-\bar{\zeta_i}\zeta_1|^2}.&
\end{eqnarray}
Take $z=0$ except
$z_{1i}={{|\bar{\xi_i}\xi_1-\bar{\zeta_i}\zeta_1|}({{\underset{i=1}{\overset{n}{\Sigma}}|\bar{\xi_i}\xi_1-\bar{\zeta_i}\zeta_1|^2}})^{1\over 2} 
}$. $z$ reaches the upper bound (\ref{above}) and verifies the norm conditions. 
Thus $d_{\tilde{D}}(\omega_{\xi},\omega_{\zeta})
=2 \sqrt{{\underset{i=2} {\overset{n}{\Sigma}}}|\bar{\xi_i}\xi_1
-\bar{\zeta_i}\zeta_1|^2}$. 
As in proposition 2, an intrinsic expression is found by
developing $\text{Tr}(\xi\xi^* -\zeta\zeta^*)^2$, and the formulas for $d_{D}$ are the same $\blacksquare$
$\\$

The cases $\mathcal{A}=M_p(\mathbb{C})\oplus M_q(\mathbb{C})$ is not studied here, neither is the space associated to a sum of three or more algebras 
with at least a noncommutative one. We focus on sums of commutative algebras. Then $\mathcal{A}$ is isomorphic to $\underset{i=1}{\overset{k}\bigoplus}\,
\mathbb{C}$. The space associated to $k=1$ has no interest.
With $k=2$, there is only one distance to compute which is equal to ${1 \over |D_{12}|}$. The generic cases $k=3,4$ with a real operator $D$, and some examples with 
$k=n\in\mathbb{N}$ are considered below. Before, we present general results on commutative finite spaces.    

\section{Commutative finite spaces.}
A n point commutative space is determined
by a triplet ($\mathcal{A},\mathcal{H},D)$ in which 
$\mathcal{A}={\overset{n}{\underset{1}{\bigoplus}}}y \, 
\mathbb{C}$ is representated over 
$\mathcal{H}= \mathbb{C}^n$  as a diagonal matrix:
$$     
\mathcal{A} \ni a
= \left(
\begin{array}{cccc}
a_1   	 & 0	 & \ldots	 	 & 0 \\
\vdots	 & a_2 &	       	& \vdots \\
\vdots  &	    & \ddots    &  \vdots   \\
0       &\ldots	&\ldots  	& a_n \\
\end{array}
\right),
$$ 
where $a_i \in \mathbb{C}$, but for distance computing we restrict to $a_i\in \mathbb{R}^+$ thanks to lemma 1.  
To make computations easier, we only consider real operators $D$. As $D$ only appears through its commutator $[D,a]$, we
assume that it has the following form: 
$$
D= \left(
\begin{array}{ccccc}
0   	 &D_{12}&\ldots&\ldots&D_{1n}	 \\
D_{12}&      &D_{23}&      &0 \\
\vdots&D_{23}&0	    &\ddots&\vdots   \\
\vdots&	     &\ddots&\ddots& D_{n-1,n}   \\
D_{1n}&\ldots&\ldots &D_{n-1,n}	& 0 \\
\end{array}
\right) \text{ with }D_{ij} \in \mathbb{R}.
$$  
Pure states can be interpreted as points of a n
point space whose function algebra is $\mathcal{A}$:
$a(i)\doteq a_i.$ 
The distance between two points $i,j$ of this finite space is 
\begin{equation}
\label{distance2}
d(i, j) = \underset {a \in \mathcal{A}^+} {\sup}  \, \{  \, |a_i - a_j|  \;\,  / \;\, \|[D,a]\| = 1 \}.
\end{equation}

In finite spaces, $D$ may be interpreted as the adjacency matrix of a lattice \cite{dimakis1}: two
points $i$ and $j$ are connected if and only if $D_{ij} \neq 0$. For instance, in the four point space, the restriction to $D_{13} = D_{24} = 0$  corresponds to a
cyclic graph: 
\begin{figure}[h]
\begin{center}
\mbox{\rotatebox{270}{\scalebox{0.45}{\includegraphics{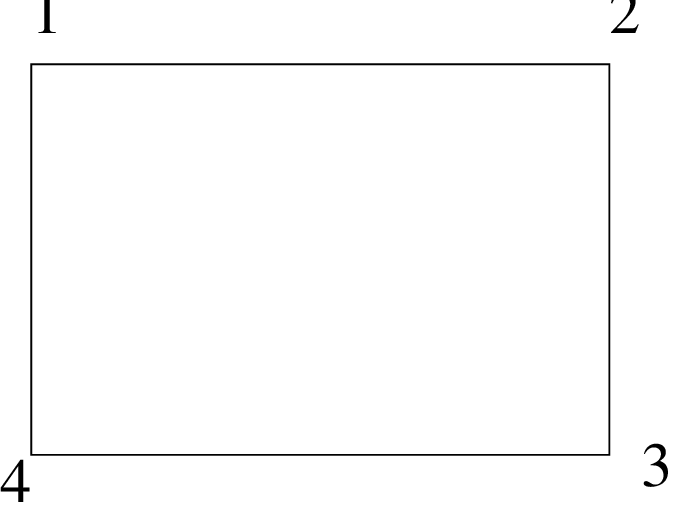}}}}
\end{center}
\end{figure}

A path $\gamma_{ij}$ is a sequence of $p$ distinct points $(i,i_2,...,i_{p-1},j)$ with $D_{i_{k}i_{k+1}}\neq 0, \, \forall k\in\{1,p-1\}$. Since $d(1,2)=
{1 \over |D_{12}|}$ in a two point space, the length of $\gamma_{ij}$ is by definition
\begin{equation*}
L(\gamma_{ij})\doteq \underset {k=1}{\overset{p-1}\Sigma}{1\over {|D_{i_ki_{k+1}}|}}.
\end{equation*}
Two points $i$,$j$ are said connected if there exists at least one path $\gamma_{ij}$.
The geodesic distance $L_{ij}$ is by definition the length of the shortest path
connecting
$i$ and
$j$.
$\\$

{ \it {\bf Proposition 4.}

{ \bf i)} Let $D'$ be the operator obtained by cancelling one or several lines and the corresponding columns of $D$. Then
$d_{D'} \geq d_{D}$.

{\bf ii)} The distance between two points $i$ and $j$ depends only on the matrix elements corresponding to  points situated on a path
joining $i$ and $j$.

{\bf iii)} The distance between two points is finite if and only if they are connected.} 
$\\$

{\it Proof.}

{\it i)} Let define $e\in\mathcal{A}$ by $e_{i}=0$ if the $i^{th}$ line and column are cancelled, $e_{i}=1$ elsewhere. $e$
is a projection, commuting with $\mathcal{A}$, and $D'=eDe$. Thus $||[D',a]||\leq ||[D,a]||, \forall a\in\mathcal{A}$. So 
$$\sup\{
|a_i-a_j|\,/\, \|[D,a]\|\leq 1 \} \geq \sup\{
|a_i-a_j|\,/\, \|[D',a]\|\leq 1 \}.$$

{\it ii)} Let $\Gamma_{ij}$ denote the graph associated to the set of points belonging to any path $\gamma_{ij}$, and
$I_{ij}$ the set of points which are not on any path
$\gamma_{ij}$. Any point of $I_{ij}$ is connected by at most one path to $\Gamma_{ij}$. In other terms, $\forall l \in I_{ij}$ there is at most one point
$m_l\in\Gamma_{ij}$ such that $l$ and $m_l$ are connected and $\gamma_{lm_l}$ has all its points (except $m_l$) in $I_{ij}$.

 Let $D'$ be the operator obtained by cancellation of all lines and columns associated to points of $I_{ij}$. Assume that the supremum in
(\ref{distance2}) with
$D'$ is reached by $a'\in \mathcal{A}=\mathbb{C}^n$. Consider $a\in\mathbb{C}^n$ such that $a_p=a'_p$ except for the points of $I_{ij}$ for which
$a_l=a_{m_l}$ or $a_l=0$ if $m_l$ does not exist. Then $||[D,a]|| = ||[D',a']||$, so $d_D(i,j) \geq d_{D'}(i,j)$. By ${\it i)}$, $d_{D}(i,j)\leq d'_D(i,j)$. Hence the result.
$\\$

{\it iii)} Suppose $i$ and $j$ are connected. There is at least one path $\gamma_{ij}=(i,i_2,...,i_{p-1},j)$ whose length is the geodesic distance
$L_{ij}$. Let obtain
$D'$ by cancelling all lines and columns which do not correspond to points of $\gamma_{ij}$. Then  $d(i,j) \leq d_{D'}(i,j)$. By the triangular
inequality, one has
$$
d_{D'}(i,j)\leq \underset{k=1}{\overset{p-1}{\Sigma}}d_{D'}(i_k,i_{k+1}) =\underset{k=1}{\overset{p-1}{\Sigma}}{1\over{|D_{i_ki_{k+1}}}|}\doteq L_{ij}.
$$
$d(i,j)$ is smaller than the geodesic distance, thus it is finite.

If $i$ and $j$ are not connected, define $a \in \mathbb{C}^n$ by $a_i=t>0, a_k=a_i$ if $k$ and $i$ are connected, $a_k=0$ otherwise. Then $[D,a]=0$
and $|a_i-a_j|=t$. As $t$ is arbitrary, $d(i,j)$ is infinite. $\blacksquare$  
$\\$

$\bullet$ For simplification purpose, we write
\begin{equation}
\label{x}
a_{ij}\doteq a_j - a_i \quad \text{ and } \quad 
x\doteq a_{21} \, , \, x_i \doteq a_{i+1,1},\quad 2\leq i \leq n-1.
\end{equation} 

In the $n=3$ and $n=4$ point case, this reduces to
\begin{equation}
\label{x4}
y\doteq a_{31} \, \text{ and } \,  z\doteq a_{41},
\end{equation}

$\bullet$ In the four point space, the components of $D$ are
\begin{equation}
\label{d4}
d_1 \doteq {1\over D_{12}},\; d_2 \doteq {1\over D_{13}},\; d_3 
\doteq {1 \over D_{14}},\; d_4 \doteq {1\over D_{23}},\; d_5 \doteq {1 \over
D_{24}},\; d_6\doteq {1 \over D_{34}}.
\end{equation}

\section{Distance on a regular space.}

A n point commutative space is called regular when all coefficients of operator $D$ are
equal: 
$$D = \{D_{ij}\} =\{ k(1 - \delta_{ij})\} \;  ,\; k \in \mathbb{R}.$$ 

{\bf Proposition 5.}

{\it {\bf i)} The distance between two points $i,j$ of a regular space of constant k is
$$ d(i,j) ={1 \over
|k|}{\sqrt{{2}
\over n}}.$$

{\bf ii)} If the link, and only this link, between two point $i_1,i_2$ is cut, $D_{i_1i_2}=0$, then 
$$ d(i_1,i_2) ={1 \over
|k|}{\sqrt{{2}
\over n-2}}.$$ }

{\it Proof.}
In the regular space, the problem is symmetrical: all distances are equal and we compute $d(1,2)$. When a link is cut, we take $i_1=1,i_2=2$ to fix notations,
and denote by ${D'}$ the operator $D$ with $D_{12}=0$ . In both case, (\ref{distance2}) and (\ref{x}) yield
\begin{equation}
\label{nptsdistance}
d(1,2) = \underset {a \in \mathcal{A}_{+}} {\sup} \{\; |x|\, /\, \|[\,D \text{ or } {D'}, a]\| =1 \}.
\end{equation}
We first compute the norm of the commutator, and then find the supremum.
$\\$

{\bf Lemma 6.}

{\it {\bf i)} 
$ \|[D,a]\|^2 =|k|^2
                         \overset{n}{\underset{i=1}{\Sigma}}
 	                        \, 
                          \overset{n}{\underset{j=i+1}{\Sigma}}  
	                         a_{ij}^{2}
             =|k|^2  \left[
                        x^2 
                        + \overset{n-1}{\underset{i=2}{\Sigma}} \left( x_i^2 + (x-x_i)^2  + \overset{n-1}{\underset{j=i+1}{\Sigma}}(x_i - x_j)^2\right)
                      \right]
                       .$

\begin{eqnarray*}
 \|[{D'},a]\|^2 &=&{|k|^2 \over 2}
                   \left[ 
                   \overset{n}{\underset{i=1}{\Sigma}}\, \overset{n}{\underset{\underset{-(1,2)}{j=i+1}}{\Sigma}}a_{ij}^2
                   +\sqrt{
                          ( \overset{n}{\underset{i=1}{\Sigma}} \,  \overset{n}{ \underset { \underset{-(1,2)}{j=i+1} } {\Sigma} }  a_{ij}^2) ^2
                          -4a_{12}^2 \overset{n}{\underset{i=3}{\Sigma}}\; \overset{n}{ \underset{j=i+1}{\Sigma}} a_{ij}^2
                          }
                   \right]\\
                     &=&{|k|^2 \over 2}\left[
                                 \overset{n-1}{\underset{i=2}{\Sigma}} \left( x_i^2 + (x-x_i)^2  + \overset{n-1}{\underset{j=i+1}{\Sigma}}(x_i -
x_j)^2\right)\right] \\
               & & + \left[ \sqrt{ 
                                          (\overset{n-1}{\underset{i=2}{\Sigma}} \left( x_i^2 + (x-x_i)^2  + \overset{n-1}{\underset{j=i+1}{\Sigma}}(x_i -
x_j)^2\right))^2 - 4x^2 \overset{n-1}{\underset{i=2}{\Sigma}}\;  \overset{n-1}{\underset{j=i+1}{\Sigma}}(x_i - x_j)^2
                                 }
                                  \right]. 
\end{eqnarray*}

{\bf ii)} For the regular space, the supremum of $x$ in (\ref{nptsdistance}) is reached when all $x_i$'s are equal.}
$\\$

{\it Proof.}
{\it i)} $C \doteq i[D,a]$ is the $n\text{ x }n$ matrix
$$
C = k\left(\begin{array}{cccc}
0 	     &ia_{12}& 	  	   &    \\
ia_{21}	&\ddots &	ia_{ij}&   \\
       	&ia_{ij}& \ddots &    \\
        &      	&       	&	 0 \\
\end{array}\right) , 
$$
with rank $\leq 2$ since its kernel is generated by the $(n-2)$ independant vectors
\begin{equation*}
\Lambda_{k} =( {{a_{k2} }\over{a_{21} }};  {{ a_{1k} } \over{ a_{21}
}}; 0;...;1;...; 0;), \; 1 \text{ beeing at the $k^{th}$ position,} \; 3\leq k\leq n. 
\end{equation*}
Moreover $C$ is hermitian and traceless, so it has two non zero real eigenvalues $\pm\lambda$. Thus $\lambda=\sqrt{{Tr(C^2)}\over 2}$. A direct
computation yields 
$$\lambda= k{\sqrt  { \overset {n}     {\underset {i=1}    {\Sigma}}\, \overset {n} {\underset {j=i+1} {\Sigma} }  
	(a_{ij})^{2}}}.$$
Finally $\|[D,a]\| =\|i[D,a]\|=|\lambda|=|k|{\sqrt  { \overset {n}     {\underset {i=1}    {\Sigma}}
 	\, \overset {n} {\underset {j=i+1} {\Sigma} }  
	(a_{ij})^{2}}}= |k|\sqrt{ x^2 +  { \overset {n-1}     {\underset {i=2}    {\Sigma}}\,  x_i^2 +  { \overset {n-1}     {\underset {i=2}    {\Sigma}}
\, \overset {n-1} {\underset {j=i+1} {\Sigma} } (x_i - x_j})^{2}}}.$

 Let ${C'} \doteq i[{D'},a]$. ${C'}$ is the $n\text{ x }n$ matrix
$$
{C'} = k\left(\begin{array}{cccc}
0 	    & 0	    &ia_{13} &  \\
0	     & 0     &        & ia_{ij} \\
ia_{31}&       &\ddots  & \\
	      &ia_{ij}&        &0   \\

\end{array}\right), 
$$
with rank $\leq 4$ since $ker({C'})$ is generated by the $(n-4)$ independant vectors
\begin{equation*}
{\Lambda'}_{p} =(0;0;{{a_{p4}}\over{a_{43}}};{{a_{3p}}\over{a_{43}}}; 0;...;1;...; 0), \; 1 \text{ beeing at the $p^{th}$ position,} \; 5\leq p\leq n. 
\end{equation*}
Beeing hermitian and traceless, ${C'}$ has four real eigenvalues $\pm{\lambda'_1}$, $\pm{\lambda'_2}$. Thus its
characteristic polynomial is
\begin{equation}
\label{carc}
\mbox{\Large$\chi$}({C'})=X^{n-4}(X^2-{\lambda'_1}^2)(X^2-{\lambda'_2}^2)
=X^n -({\lambda'_1}^2 + {\lambda'_2}^2)X^{n-2}
+{\lambda'_1}^2{\lambda'_2}^2X^{n-4}.
\end{equation} 
A direct computation yields ${\lambda'_1}^2 + {\lambda'_2}^2={1 \over 2}Tr({C'}^2)= k^2{\overset{n}{\underset{i=1}{\Sigma}}\,\overset{n}{\underset{\underset{-(1,2)}{j=i+1}} 
{\Sigma}}(a_{ij})^{2}}.\\$
The coefficient of $X^{n-4}$ is the sum of all the minors of ${C'}$ of degree $4$. A minor $M(1,k,l,p)$
composed with the first (or second column) and three others columns $k,l,p \notin \{1,2\}$ (and the associated lines) is also a minor
of $C$. As $C$ is of rank $\leq 2$, its minors of degree greater than 2 are null, so $M(1,k,l,p)=M(2,k,l,p)=0$. The same is true for the minors
$M(q,k,l,p)$ with $q\notin \{1,2\}$. Finally, the only non zero minors are 
$$M(1,2,l,p)= k^4 Det\left(
\begin{array}{cccc}
0     & 0      & ia_{1l} & ia_{1p} \\
0     & 0      & ia_{2l} & ia_{2p} \\
ia_{l1}& ia_{l2} & 0      & ia_{lp}\\
ia_{p1}& ia_{p2} & a_{pl} & 0
\end{array}
\right)
=k^4 Det\left(
\begin{array}{cc}
a_{1l} & a_{1p} \\
a_{2l} & a_{2p}
\end{array}
\right)^2=k^4 a_{21}^2 a_{pl}^2.$$
Summing all these minors gives
${\lambda'_1}^2{\lambda'_2}^2 =a_{12}^2 \underset{l=3}{\overset{n}\Sigma}\; 
\underset{p=l+1}{\overset{n}\Sigma}y a_{pl}^2.$
Then, solving (\ref{carc}) yields
$$\|[{D'},a]\|^2 ={|k|^2 \over 2}
                   \left( 
                   \overset{n}{\underset{i=1}{\Sigma}}\;  \overset{n}{\underset{\underset{-(1,2)}{j=i+1}}{\Sigma}}a_{ij}^2
                   +\sqrt{
                          ( \overset{n}{\underset{i=1}{\Sigma}} \,  \overset{n}{ \underset { \underset{-(1,2)}{j=i+1} } {\Sigma} }  a_{ij}^2) ^2
                          -4a_{12}^2 \overset{n}{\underset{i=3}{\Sigma}}\;  \overset{n}{ \underset{j=i+1}{\Sigma}} a_{ij}^2
                          }
                   \right).$$

{\it ii)} Let $f(x,x_2,...,x_{n-1})\doteq x^2+{ \overset{n-1}{\underset{i=2}{\Sigma}}x_i^2 + (x-x_i)^2 +{\overset {n-1}{\underset {i=2}{\Sigma}}
\overset {n-1}{\underset{j=i+1}{\Sigma}}(x_i-x_j})^{2}}$ and suppose that $(x,x_2,...,x_{n-1}) \in \mathbb{R}^{n-1}$ reaches the supremum, namely 
$$
f(x,x_2,...,x_{n-1})={1\over|k|^2} \; \text{ and } \; d(1,2) = |x|,
$$then 

{\it a) $x$ is positive:} from the global parity of $f$, $x$ can be chosen positive. 

{\it b) $x_i\leq {{x}\over{2}},\; \forall i \in \{2,...,n-1\}:$}
suppose that $p$ of the $x_i$'s are greater than $x \over 2$ and denote them generically by $x_p$. Consider now the (n-1)-uplet in which
all $x_p$'s are replaced by ${x\over 2}$. Then  
$f$ decreases for $x_p^2 + (x-x_p)^2 \geq {{x^2}\over 4} + (x-{x\over 2})^2$ and $(x_i - x_p)^2  \geq (x_i - {x\over 2})^2$. Fixing the values of the remaining $x_i$'s leads to see $f$ as a function of the single variable
$x$: 
\begin{eqnarray*}
&f(x)= x^2 + p({x\over 2})^2 +
{\underset {i}{\Sigma}}   x_i^2             + p(x-{x\over 2})^2 +  {\underset {i}{\Sigma}}(x-x_i)^2+
{\underset {i}{\Sigma}}p({x\over 2}- x_i)^2 + 
{\underset {i}{\Sigma}}
{\underset {j}{\Sigma}}(x_j - x_i)^2,& \\
&f'(x)= 2x + 2px + 
2{\underset {i}{\Sigma}}(x-x_i) +
{\underset {i}{\Sigma}}p({x\over 2}-x_i).&
\end{eqnarray*}
As $x_i \leq {x \over 2} \leq x$ , $f'(x) > 0$ when $x > 0$. $f$ is continue and ${\underset {x\rightarrow \infty}{\lim}}f(x) = +\infty$, 
 so there is $x_0 > x$ such that $f(x_0)={1\over|k|^2}$. In other terms, the initial (n-1)-uplet in $x_p$ does not reach the supremum which is in contradiction with
our hypothesis. So
$p=0$.

{\it c)  $x_i \geq 0 ,\; \forall i \in \{2,...,n-1\}$}: the proof is the same by replacing all  $x_i \leq 0$ by ${x \over 2}$.

{\it d) All $x_i$'s are equal}:
 let $\lambda$ and $\Lambda$ be the two smallest value of the $x_i$'s, with $\lambda \leq \Lambda$ . If $\lambda=\Lambda$ then it comes
immediatly that all
$x_i$'s are equal. If $\lambda\neq\Lambda$ then 
$$0 \leq \lambda < \Lambda \leq x_i \leq {x
\over 2} \; ,\; \forall i \in \{2,...,n-1\}.$$ 
Assume that $m$ of the $x_i$'s are equal to $\lambda$.
Summing over $x_i \neq \lambda$, one obtains:
\begin{eqnarray*}
f(x,x_2,...,x_{n-1}) = x^2 + m\lambda^2 + {\underset {i} \Sigma} x_i^2 + {\underset {i} \Sigma} (x-x_i)^2+ m(x-\lambda)^2 + {\underset {i}\Sigma}
m(\lambda -x_i)^2 + {\underset {i,j}\Sigma} (x_i -x_j)^2.
\end{eqnarray*}
Fix the values of $x_i \neq \lambda$ and consider now $\lambda$ not like a constant but like the value of a variable
$x_{min}$. Then $f$ can be seen as a function $f_m$ of the two variables $x_{min}$ and $x$ with 
$${{\partial f_m} \over {\partial x_{min}}}(x_{min},x) = 2mx_{min} + 2m(x_{min} - x) + {\underset {i} \Sigma} 2m(x_{min} -x_i).$$
As ${{\partial f_m} \over {\partial x_{min}}}(x_{min},x)< 0$ for $x_{min} \in [\lambda,\Lambda[$, one has $
f_m(\Lambda,x) < f_m(\lambda,x)={1\over|k|^2}.$
Moreover
$${{\partial f_m} \over {\partial x}}(\Lambda,x) = 2x + 2m(x-\Lambda) + {\underset {i} \Sigma} 2(x-x_i) > 0,$$
so there is  $x_0 > x$ such that
$f_m(x_0,\Lambda)={1\over|k|^2}$, which is inconsistent with our hypothesis. So $\lambda = \Lambda$.$\blacksquare$
\\

{\bf Proof of proposition 5.}
{\it i)} According to lemma 5, $x_i = x_2, \; 2\leq i \leq n-1$. The norm condition in (\ref{nptsdistance}) becomes
 $$ 2(n-2)x_2^2 + [2(2-n)x]x_2 + [(n-1)x^2 - {1\over {|k|^2}}] = 0,\\$$
which has no real solutions in $x_2$ unless $|x|\leq {1\over |k|}\sqrt{{2}\over {n}}$. This upper bound is reached when 
$x_2={x\over 2} ={1\over 2|k|}\sqrt{{2 \over n}}$.

{\it ii)} Let $\; h_1(x,x_i) \doteq \overset{n-1}{\underset{i=2}{\Sigma}} x_i^2+(x-x_i)^2,\; 
h_0(x_i)   \doteq \overset{n-1}{\underset{i=2}{\Sigma}}\, \overset{n-1}{\underset{j=i+1}{\Sigma}}(x_i - x_j)^2,\; g(x,x_i)\doteq h_1(x,x_i)-2x^2$. Lemma
6 yields
\begin{equation}
\label{h}
\|[{D'},a]\|^2={{|k|^2}\over 2}\left(  h_1 +h_0 +\sqrt{h_1^2 + h_0^2 + 2g.h_0}  \right) .
\end{equation}
Let $x_0=\underset{x,x_i \in \mathbb{R}}{\sup} \{ x / h_1(x,x_i)= {1\over|k|^2} \}$. As $g$ and $h_0$ are both positive, (\ref{h}) implies that $d(1,2)\leq x_0$.
Imitating $i)$, one finds this upper bound is reached when all $x_i$'s are equal, and $x_0={1\over |k|}\sqrt{{2}\over{n-2}}$. $\blacksquare$
$\\$

In finite spaces which are not regular, distances are not always explicitly computable. The cases $n=3$ and $n=4$ are considered below. 

\section{Three point space.}

{\bf Proposition 7.}
{\it For a three point space with operator
$$
D = \left(
\begin{array}{ccc}
0 & D_{12} & D_{13}\\
D_{12} & 0 & D_{23}\\
D_{13} & D_{23} & 0
\end{array}
\right) \; D_{ij}\in \mathbb{R}, \quad
d(1,2)=\sqrt{{D_{13}^2+D_{23}^2}\over{D_{12}^{2}D_{13}^{2}+D_{12}^{2}D_{23}^{2}+D_{23}^{2}D_{13}^{2}}}.\\
$$
the others distances come from suitable permutations of indices.}
$\\$

{\it Proof.}
Equation (\ref{distance2}) and notations (\ref{x4}) gives
$$
d(1,2) = \underset {a \in \mathcal{A}_{+}}{\sup} \,\{x\;/\; \|[D,a]\|=\left\| \left(
\begin{array}{ccc}
0 & -D_{12}x & -D_{13}y\\
D_{12}x & 0 & D_{23}(x-y)\\
D_{13}y & D_{23}(y-x) & 0
\end{array}
\right) \right\|=1\}.
$$
By direct calculation,
 $\, \|[D,a]\|=\sqrt{ D_{23}(x-y)^2+D_{13}y^{2}+D_{12}x^{2} }.\, $
Thus $d(1,2)$ is the largest value of $x$ for which there is a point $(x,y)$ belonging to the ellipse 
\begin{equation}
\label{ellipse}
(D_{23}^{2}+D_{12}^{2})x^{2}+(D_{13}^2+D_{23}^{2})y^2-2D_{23}^2xy=1.
\end{equation}
$d(1,2)$ is the positive $x$ for which the equation in $y$ (\ref{ellipse}) has a zero discriminant, that is
for
$$
x=\sqrt{{D_{13}^2+D_{23}^2}\over{D_{12}^{2}D_{13}^{2}+D_{12}^{2}D_{23}^{2}+D_{23}^{2}D_{13}^{2}}}. \blacksquare$$

The three distances verify an inequality of the triangle "power two" since
\begin{equation}
\label{trgcarre}
d(1,2)^2 + d(2,3)^2 \geq  d(1,3)^2,
\end{equation}
and two others inequality by permutations. The question rises of inverting the problem: for three positive numbers $(a,b,c)$ verifying 
(\ref{trgcarre}), is there any operator $D$ giving $(a,b,c)$ as noncommutative distances ?
\bigskip

{\bf Proposition 8.} 
{\it Let $a,b,c \in \mathbb{R}^+$ verifying $a^2+b^2\geq c^2,\, b^2+c^2 \geq a^2, \, a^2+c^2 \geq b^2.$
Then, there is an operator $D$ such that $d(1,2)= a,\, d(1,3)= b,\, d(2,3)=c,$ explicitly given by
$$
D_{12}=\sqrt{{{2(b^2 + c^2 - a^2)}\over{(a+b+c)(-a+b+c)(a-b+c)(a+b-c)}}},$$
$D_{13}$ and $D_{23}$ coming from permutations of $a,b,c$.}
\bigskip

{\it Proof.}
Writing ${1\over{D_{12}^2}}=R_{12},\,{1\over{D_{23}^2}}=R_{23},\,{1\over{D_{13}^2}}=R_{13},$ proposition 7 gives
\begin{equation*}
{1\over{d(1,2)^2}}={1\over{R_{12}}}+{1\over{R_{23}+R_{13}}}.
\end{equation*}
 $d(1,2)^2 $ is the electrical resistance between points 1 and 2 of the triangle circuit. 
\begin{figure}[h]
\begin{center}
\mbox{\rotatebox{0}{\scalebox{0.7}{\includegraphics{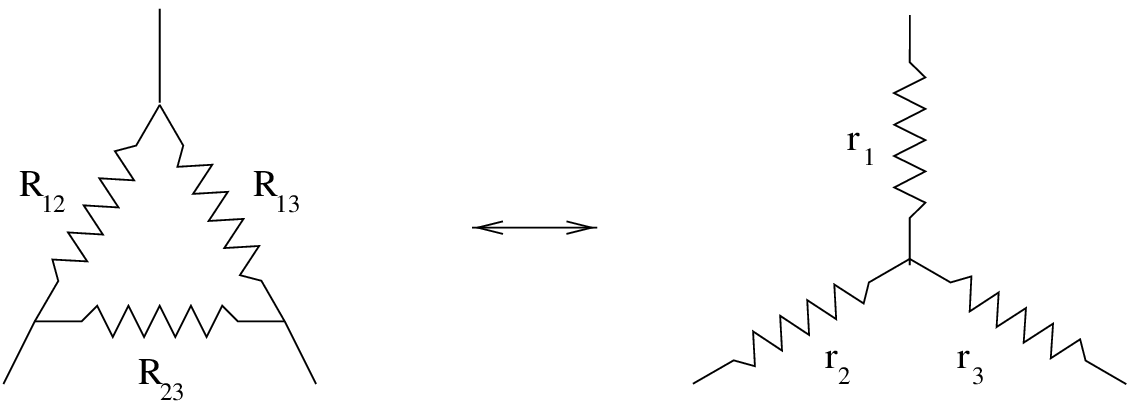}}}}
\end{center}
\end{figure} 
Finding $D_{ij}$ means finding three elements $R_{kp}$ that induce a resistance $d(i,j)^2$ between points $i,j$. A classical result \cite{ed} 
indicates that the triangle circuit is equivalent to a stellar circuit $(r_1,r_2,r_3)$ with
\begin{equation}
\label{resistance}
R_{12}={1\over r_3}({r_1r_2 + r_1r_3 + r_2r_3}),
\end{equation}
$R_{13}$ and $R_{23}$ coming from suitable permutations of indices. In the stellar circuit
\begin{equation*}
\left\{
             \begin{array}{c}
             d(1,2)^2= r_1 + r_2 \\
             d(1,3)^2= r_1 + r_3 \\
             d(2,3)^2= r_2 + r_3 
             \end{array}
\right.
\text{ which implies }
\left\{
             \begin{array}{c}
             2r_1=d(1,2)^2 + d(1,3)^2 - d(2,3)^2 \\
             2r_2=d(1,2)^2 + d(2,3)^2 - d(1,3)^2 \\
             2r_3=d(1,3)^2 + d(2,3)^2 - d(1,2)^2.
             \end{array}
\right. 
\end{equation*}
Inserting in (\ref{resistance}) this gives
 $$D_{12}=\sqrt{{{2(d(1,3)^2+d(2,3)^2-d(1,2)^2)}\over{2 (d(1,2)^2d(1,3)^2
+d(1,2)^2d(2,3)^2+d(1,3)^2d(2,3)^2)-d(2,3)^4-d(1,3)^4-d(1,2)^4}}}.\blacksquare$$

\section{Four point space.}

In a three point space, distances can be computed for any real operator $D$. This is not true for a four point space. Notations (\ref{x4}) and (\ref{d4}) are used.
$\\$

{\bf Theorem 9.}

{\it {\bf i)} $d(1,2)$ is a root of a polynomial of degree $\delta\leq 12$.

{\bf ii)} $d(1,2)$ is generically not computable: it is a root of a polynomial not solvable by radicals. 

{\bf iii)} It is computable in the following case: when ${1 \over {d_2}}={1 \over {d_5}}=\infty$,
\begin{eqnarray*}
d(1,2)&=& \left\{ 
                \begin{array}{ll}
                  d_1&\text{if }\quad d_1^2\leq d_6^2\text{, else}\\
                  {\frac{{d_1}\,{\sqrt{{{\left( {{{d_3}}^2} + {d_1}\,{d_6} \right) }^2}}}}{ {\sqrt{{{{d_1}}^2} + {{{d_3}}^2}}}\,{\sqrt{{{{d_3}}^2} +
                  {{{d_6}}^2}}}}} &\text{if } \quad d_1d_6 = d_3d_4\text{, else}\\
                  \sqrt{{ d_1^2(d_3^2 + d_6^2)(d_4^2 + d_6 ^2)}\over {(d_3d_4 - d_1d_6)^2}}&\text{if } C \leq 0\text{, else}\\
                  \max(\sqrt{{ d_1^2(d_3^2 + d_4^2)}\over{(d_3+d_4)^2 + (d_1-d_6)^2}},\sqrt{ { d_1^2(d_3^2 - d_4^2)}\over{(d_3-d_4)^2 +(d_1+d_6)^2}}),&\\
                 \end{array} \right. \\
\text{where } C &=& ({{({d_3} + {d_4}) }^2}{d_6} 
+ ({d_1} - {d_6})({d_3}\,{d_4} - {{{d_6}}^2})) 
({{( {d_3} - {d_4} ) }^2}{d_6} +({d_1}+{d_6}) \,( {d_3}{d_4} +
{{{d_6}}^2})).
\end{eqnarray*}

$$
d(1,3)= \left\{ \begin{array}{ll} 
        {\sqrt{{{{d_3}}^2} + {{{d_6}}^2}}} 
&\text{if }(d_3^2 + d_6^2) \leq (d_1d_6 -d_3d_4)^2,\\
        {\sqrt{{{{d_1}}^2} + {{{d_4}}^2}}}&
\text{if }(d_1^2 + d_4^2) \leq (d_1d_6 - d_3d_4)^2,\\
       \max({\frac{{\sqrt{{{\left( {d_1}\,{d_3} +{d_4}\,{d_6} \right) 
       }^2}}}}{{\sqrt{{{\left( {d_3} + {d_4}\right) }^2} + 
            {{\left( {d_1} - {d_6} \right)}^2}}}}},
{\frac{{\sqrt{{{\left( {d_1}\,{d_3} +{d_4}\,{d_6} \right) 
}^2}}}}{{\sqrt{{{\left( {d_3} - {d_4}\right) }^2} + 
                                 {{( {d_1} + {d_6} )^2}}}}}} )&
\text{otherwise}.
                \end{array} \right.$$

Permuations of the $d_i's$ give $d(2,3), d(3,4), d(1,4)$ (resp. $d(2,4)$) from $d(1,2)$ (resp. $d(1,3)$).}
$\\$
The rest of this section is devoted to the proof of theorem 9. 
$\\$

$\bullet \;  
[D,a] = \left(
\begin{array}{cccc}
0 	&  -{x \over d_1} 	 &-{y\over d_2}	 & -{z \over d_3}\\
x \over d_1	& 0 &	 {{x-y} \over d_4}	&{{x-z} \over d_5} \\
	y \over d_2 &{y-x} \over d_4	& 0 &{{y-z} \over d_6}	 \\
 z \over d_3      	&	{{z-x} \over d_5}&z-y \over d_6	& 0    
\end{array} \right)
\;
 \text{ and } \overrightarrow{r_a} \doteq \left(
\begin{array}{c} x\\ y\\z \end{array}
\right), \; \forall a \in \mathcal{A}_{+}= \mathbb{R^+}^4.$
$\\$

$\bullet$ For $\overrightarrow{r}\doteq (x,y,z) \in \mathbb{R}^3$, define the functions:
\begin{eqnarray}
\nonumber
\alpha (\overrightarrow{r}) &\doteq& {x^2 \over {d_1}^2}  +{y^2 \over {d_2}^2} + {z^2 \over {d_3}^2} + {(x-y)^2 \over {d_4}^2} +{(x-z)^2 \over
{d_5}^2} + {(y-z)^2
\over {d_6}^2},\\
\nonumber
\beta({\overrightarrow r}) &\doteq&{{ x(y-z)} \over {d_1 d_6}} + { { z(x-y)} \over {d_3 d_4}} 
+{ { y(z-x)} \over {d_2 d_5}},\\
\nonumber
n(\overrightarrow{r}) &\doteq& \alpha(\overrightarrow{r}) + \sqrt{\alpha(\overrightarrow{r})^2-4\beta(\overrightarrow{r})^2},\\
\nonumber
f(\overrightarrow{r}) &\doteq& \alpha(\overrightarrow{r}) - \beta(\overrightarrow{r})^2 -1,
\end{eqnarray}
and the surfaces $\mathcal{N}$ and $\mathcal{F}$: 
\begin{eqnarray*}
\mathcal{N} &\doteq& \{\overrightarrow{r}\in
\mathbb{R}^3 \; / \; n(\overrightarrow{r}) = 2\},\\
\mathcal{F}&\doteq& \{ \overrightarrow{r} \in \mathbb{R}^3 \; / \; f(\overrightarrow{r})=0 \},\quad \text{ with } \mathcal{N}\subset \mathcal{F}.
\end{eqnarray*}

{\bf Lemma 10.}

{\it {\bf i)} For $a \in \mathcal{A}_{+}, \,  \|[D,a]\|^2= {1 \over 2}n(\overrightarrow{r_a}).$}

{\bf ii)}
{\it For $\overrightarrow{r} \in \mathcal{N}$ such that }$\alpha(\overrightarrow{r})=2, \; \overrightarrow{\text{grad}}(f)(\overrightarrow{r}) = 0.$
$\\$

{\it Proof.} 
{\it i)} The four eigenvalues of $i[D,a]$ are
$
\lambda_i = \pm {1 \over \sqrt{2}} \sqrt{\alpha \pm \sqrt{ \alpha ^2 - 4 \beta ^2}}(\overrightarrow{r_a}),
$ so $$\|[D,a]\|^2={1 \over 2}(\alpha + \sqrt{ \alpha ^2 - 4 \beta ^2})(\overrightarrow{r_a}).$$

{\it ii)}
We show that ${{\partial f} \over {\partial y}} (\overrightarrow{r}) = 0$, the proof beeing the same for the other components of
$\overrightarrow{\text{grad}}(f)$. 

As $\overrightarrow{{r}} \in \mathcal{N}\subset \mathcal{F}$ and $\, \alpha(\overrightarrow{r}) =2$,\;  $\beta(\overrightarrow{r})
=\pm 1$. If $\; \beta(\overrightarrow{r}) =  1$ then 
$$
\alpha(\overrightarrow{{r}}) = 2 \beta(\overrightarrow{{r}}) \;
\text{ and }\;  
{{\partial f} \over {\partial y}}(\overrightarrow{r}) = {{\partial \alpha} \over {\partial y}}(\overrightarrow{r})
-2\beta(\overrightarrow{{r}}){{\partial \beta} \over {\partial y}}(\overrightarrow{r}) = {{\partial \alpha} \over {\partial
y}}(\overrightarrow{r}) - 2{{\partial \beta} \over {\partial y}}(\overrightarrow{r}).$$
Explicit calculation of
$\alpha(\overrightarrow{r}) - 2 \beta(\overrightarrow{r})=0$ shows that 
$${x \over {d_1}} = {{y-z} \over {d_6}},\; {y \over {d_2}} = {{z-x} \over {d_5}},\; {z \over{d_3}} = {{x-y}\over {d_4}},$$ 
which leads to ${{\partial \alpha} \over {\partial y}}(\overrightarrow{r}) - 2{{\partial \beta} \over {\partial y}}(\overrightarrow{r})=0$.
The proof is the same if assuming $\beta(\overrightarrow{r})=-1$.$\blacksquare$
$\\$

Thanks to notations and lemma above, (\ref{distance2}) leads to  
\begin{equation}
\label{rajout}
d(1,2) = \sup \;\{ \langle 1|{\overrightarrow{r_a}}|1\rangle\; /\; {\overrightarrow {r_a}}\in\mathcal{ N}\}.
\end{equation}
This formula is not useful, for $\mathcal{N}$ is not defined by a quadratic form. It is easier to work with $\mathcal{F}$.
$\\$

{\bf Proposition 11.}
${\it d(1,2) \in  \{ \langle 1|{\overrightarrow{r}} |1\rangle \; / \; {\overrightarrow {r}} \in \mathcal{F} \text{ and } {{\partial f \over {\partial
y}}(\overrightarrow{r})}={{\partial f \over {\partial z}}(\overrightarrow{r})}=0\}.
}$
\bigskip

{\it Proof.}
The supremum in (\ref{rajout}) is reached at a point $\overrightarrow{r}$ such that $\overrightarrow{\text{grad}}(n)(\overrightarrow{r})$, if it is defined, is parallel to
the $x$ axis. If $\alpha(\overrightarrow{r})=2$, then
$\overrightarrow{\text{grad}}(n)(\overrightarrow{r})$ is not defined but ${{\partial f} \over {\partial y}}(\overrightarrow{r})={{\partial f} \over
{\partial z}}(\overrightarrow{r})=0$ by lemma 10.  If $\alpha(\overrightarrow{r})\neq 2$, then $\overrightarrow{\text{grad}}(f)(\overrightarrow{r})$ is
colinear to $\overrightarrow{\text{grad}}(n)(\overrightarrow{r})$, so ${{\partial f}\over {\partial y}}(\overrightarrow{r})={{\partial f}
\over {\partial z}}(\overrightarrow{r})=0.$ 
To complete the proof, one just remarks that $\forall \overrightarrow{r} \in \mathbb{R}^3$, there is $a \in
\mathcal{A_+}$ such that $\overrightarrow{r} = \overrightarrow{r_a}$, for instance $a=(\xi, \xi-x,\xi-y,\xi - z)$ where $\xi \doteq \sup\{|x|,|y|,|z|\}.\blacksquare$
$\\$

According to this proposition, the distance is a common root of a polynomial in several variables and its various derivatives. Before undertaking
explicit calculations, we recall general results about polynomial systems. 
$\\$

{\bf Notes on systems of polynomial equations.}

Let $P$ and $Q$ be two polynomials of the form
\begin{eqnarray*}
P(x) &=& a_n x^n + a_{n-1} x^{n-1} + ... + a_0\\
Q(x) &=& b_m x^m + b_{m-1} x^{m-1} + ... + b_0.
\end{eqnarray*}
with $a_n,b_m \neq 0$. Without calculating the roots $p_i$, $q_j$ of $P$, $Q$, one finds by algebraic manipulations \cite{lang} of the coefficients $a_i$ and $b_j$
the resultant of $P$ and $Q$:
\begin{equation}
\label{resultante}
Res(P,Q)\doteq a_n^m b_m^n\,{\underset {i,j}{\Pi}} (p_i - q_j), \quad1\leq i \leq n, \; 1 \leq j \leq m.
\end{equation}
$Res(P,Q)$ is a polynomial in the $a_i$'s and $b_j$'s. $P$ and $Q$ have a common root if  and only if their resultant is zero. 
A particular
resultant is the discriminant:
\begin{equation*}
Dis(P)\doteq Res(P,P').
\end{equation*}
$P$ has a double root if and only if $Dis(P)= 0$.
If $P$ and $Q$ are polynomials in $x,y,z$, then $Res[P,Q,y]$ denotes the resultant of $P$ and $Q$ seen as polynomials in
$y$. Equivalently $Dis[P,z]$ stands for discriminant of $P$ seen as a polynomial in $z$. 
$\\$

{\bf Proposition 12.}
{\it Let $P(x,y,z)$ be a polynomial of degree 2 in $z$, whose coefficients are real functions of $x$ and $y$. If $P(x_0,y_0,z_0)={{\partial P}\over{\partial
y}}({x_0,y_0,z_0}) = {{\partial P}\over{\partial z}}({x_0,y_0,z_0}) =0$ for some $(x_0,y_0,z_0) \in \mathbb{R}^3$, then $x_0$ is a root of the polynomial
$Dis[Dis(P,z),y]$, and $y_0$ is a double root of the polynomial $Dis(P,z)(x_0,y)$.} 
\bigskip

{\it Proof.} Writing $P(x,y,z) = a(x,y)z^2 + b(x,y)z + c(x,y)$, a direct computation yields
\begin{eqnarray*}
V                                &\doteq  & Dis(P,z) = a(4ac - b^2),\\
{{\partial V} \over {\partial y}}&=       & {{\partial a}\over{\partial y}}(8ac - b^2)- 2ab{{\partial b}\over{\partial y}} + 4a^2{{\partial c}\over{\partial y}},\\
Res({{\partial P}\over{\partial y}},{{\partial P}\over{\partial z}},z)&=& b^2{{\partial a}\over{\partial y}} -2ab{{\partial b}\over{\partial y}} + 
4a^2{{\partial c}\over{\partial y}}.\end{eqnarray*}
$P(x_0,y_0,z_0)= {{\partial P}\over{\partial z}}({x_0,y_0,z_0})$ implies $V(x_0,y_0)=0$, i.e. $(8ac-b^2)(x_0,y_0)=b^2(x_0,y_0)$, thus
\begin{equation*}
Res({{\partial P}\over{\partial y}},{{\partial P}\over{\partial z}},z)({x_0,y_0})={{\partial V} \over {\partial y}}§({x_0,y_0}).
\end{equation*}
Therefore, ${{\partial P}\over{\partial y}}({x_0,y_0,z_0})={{\partial P}\over{\partial z}}({x_0,y_0,z_0})$ implies
${{\partial V}\over {\partial y}}({x_0,y_0})=0 = V(x_0,y_0),$ 
thus $y_0$ is a double root of $V(x_0,y)$ and
$$Dis(V,y)(x_0)=Dis[Dis[P,z],y](x_0)=0.\; \blacksquare$$
$\\$

 {\bf Proof of theorem 9.} Proposition 11 and 12 yields
$$d(1,2) \in  \{  x \,  / \, Dis[V(x,y),y]=0\} \text{ with } V(x,y)=Dis[f(x,y,z),z].$$
Instead of $V(x,y)$, one uses the effective form without the correctif term $a_n^m b_m^n$
appearing in (\ref{resultante}), so that zeros of
$V_{eff}$ correspond exactly to the existence of a common roots of $f$ and ${\partial f} \over {\partial y}$:
$$V_{eff}(x,y) \doteq \text{Numerator}({{Dis(f,z)}\over{n^nf_n^{2n-1}}} ),$$ 
with $f_n$ the leading coefficient of $f$ seen as a polynomial in $z$
and $n=\text{deg}(f)$. Note that the numerator is taken after a possible (but not always possible) simplification of the fraction.
$\\$

{\it  i)} By direct computation, $V_{eff}(x,y)=V_i\,y^i, \; 0\leq i\leq 4$. Exact expressions of the $V_i$'s are given in appendix. They are polynomial in $x$ of the form:
\begin{equation*}
V_4(x)=v_{4_0},\,V_3(x)=v_{3_1}x,\,V_2(x)=v_{2_2}x^2+v_{2_0},\,V_1(x)=v_{1_3}x^3+v_{1_1}x,\,V_0(x)=v_{0_4}x^4+v_{0_2}x^2+v_{0_0}.
\end{equation*}
 The discriminant $J$ of a polynomial $C=C_i y^i$ of degree four is
\begin{eqnarray*}
J(C)&\doteq& Res[C,C']= {C_4}({{{C_3}}^2}({{{C_1}}^2}{{{C_2}}^2} - 4{{{C_1}}^3}{C_3}+ 18{C_0}{C_1}{C_2}{C_3} - {C_0}( 4{{{C_2}}^3} + 27{C_0}{{{C_3}}^2}
))\\
  &&+ 2( -2{{{C_2}}^3}({{{C_1}}^2} - 4{C_0}{C_2} ) + {C_1}{C_2}( 9{{{C_1}}^2} - 40{C_0}\,{C_2} ){C_3} 
- 3{C_0}( {{{C_1}}^2} - 24{C_0}{C_2} ){{{C_3}}^2} ){C_4}\\
&&-( 27{{{C_1}}^4} - 144{C_0}{{{C_1}}^2}{C_2} + 128{{{C_0}}^2}{{{C_2}}^2} + 
       192{{{C_0}}^2}{C_1}{C_3}){{{C_4}}^2} + 256{{{C_0}}^3}{{{C_4}}^3}).
\end{eqnarray*}
Replacing $C_i$ by $V_i(x)$ shows that $J$ is
an even polynomial in $x$ of degree $\delta\leq 12$. 
$\\$

{\it ii)} We compute an explicit counterexample: assume $d_1 = d_2 = d_3 = d_4 = d_6 = 1$ and ${1 \over d_5}=0$. Then 
$$f(x,y,z)= x^2 + y^2 + z^2 + (x-y)^2 + (y-z)^2 -(x(y-z) +
z(x-y))^2.$$ 
It is a polynomial of degree 2 in the variables $x,y$ and $z$. Direct computation gives
\begin{eqnarray*}
V_{eff}(x,y) &=& 2 - 6\,{y^2} + 3\,{y^4} + 
 4\,{x^2}\,\left( -1 + {y^2} \right)  - 
  4\,x\,y\,\left( -1 + {y^2} \right),\\
Dis(V_{eff},y) &=& -768\,\left( -54 - 54\,{x^2} + 135\,{x^4} + 
    296\,{x^6} - 368\,{x^8} + 128\,{x^{10}}
    \right).
\end{eqnarray*}
Let $p(x)=-128 x^5 + 368x^4 - 296 x^3 - 135x^2 +
54x + 54$. $p$ has one real root $x_1$, two distinct complex ones $x_2$ \& $x_4$ and their conjugates $x_3$ \& $x_5$. Galois theory showes that $p$ 
is not solvable by radical (for a comprehensive presentation see Ref.\ \citelow{galois}):

$p$ is irreductible over $\mathbb{Z}$ because it is irreductible over $\mathbb{Z}_5$. Indeed modulo 5 it becomes
$q(x) = 2(x^5 + 4x^4 + 2x^3 +2x +2)$ which has no roots in $\mathbb{Z}_5$.
Therefore $p$ is irreductible over $\mathbb{Q}$.

Let $E/\mathbb{Q}$ be a splitting field extension of $p$.
As $p$ has five distinct roots, its Galois group  $\mathbb{G}=Gal(E/\mathbb{Q})$ is isomorphic to a subgroup of the symmetry group
$S_5$ which is the permutation group of $X\doteq\{x_1,x_2,x_3,x_4,x_5\}$.
As $p$ has no repeated roots, $p$ is separable so
$|\mathbb{G}|=[E/\mathbb{Q}]$ where $|\mathbb{G}|$ denotes the order of $\mathbb{G}$ and $[E/\mathbb{Q}]$ is the index, i.e. the number of cosets $\mathbb{Q}$ in
$\mathbb{G}$. 
If $\alpha$ is a root of $p$ then $[\mathbb{Q}(\alpha),\mathbb{Q}]=5$ so
$|\mathbb{G}|=[E/\mathbb{Q}]=[E/\mathbb{Q}(\alpha)][\mathbb{Q}(\alpha),\mathbb{Q}]$ is divisible by 5. Thus  $\mathbb{G}$ contains an
element of order 5: the 5-cycle $\tau=(12345)$.

The restriction to $X$ of the complex conjugation gives rise to an element $\sigma$ of $\mathbb{G}$: $\sigma=(23)(45)$. As
$\sigma$ is of order two,
$|\mathbb{G}|$ is divisible by 2. Moreover $\tau\sigma = (124)\in\mathbb{G}$ is of order three which divides $|\mathbb{G}|$. Thus $|\mathbb{G}|$ is a multiple of
$5\times 2\times 3=30$ and divides $|S_5|=120$. Since $S_5$ has no subgroup  of order 30, $|\mathbb{G}|\in \{60,120\}.$  If $|\mathbb{G}|=60$ then
$\mathbb{G}=A_5$ but $\tau \notin A_5$. So $\mathbb{G}=S_5$.

$S_n$ is solvable for $n\leq 4$ but is not solvable for $n \geq 5$,
so $\mathbb{G}$ is not solvable. Then, by Galois theorem, $p$ is not solvable by radicals.
$\\$

{\it iii)}
When $d_2 = d_5 = 0$ and $d_1d_6\neq d_3d_4$:
\begin{eqnarray*}
Dis(V_{eff})=-16d_1^{16}d_3^{14}d_4^{12}d_6^{14}(d_4^2+d_6^2)(x^2-d_1^2)(x^2(d_3d_4-d_1d_6)^2-d_1^2(d_3^2+d_6^2)(d_4^2+d_6^2))\\
             (x^2((d_3-d_4)^2+(d_1+d_6)^2)-d_1^2(d_3-d_4)^2)^2(x^2((d_3+d_4)^2+(d_1-d_6)^2)-d_1^2(d_3+d_4)^2)^2.
\end{eqnarray*}
This polynomial has four single roots $\pm x_0 , \pm x_1$ and four double roots $\pm x_2 , \pm x_3$:
\begin{eqnarray*}
&x_0=|d_1|,\,x_1=|d_1|{{\sqrt{(d_3^2+d_6^2)(d_4^2+d_6^2)}}\over{|d_3d_4 - d_1d_6|}},&\\
&x_2=|d_1|\sqrt{ {{(d_3^2 + d_4^2)}\over{(d_3+d_4)^2 + (d_1-d_6)^2}}},\, x_3=|d_1|\sqrt{ {{(d_3^2 - d_4^2)}\over{(d_3-d_4)^2+(d_1+d_6)^2}}}.&
\end{eqnarray*}
By proposition 11 and 12, $d(1,2)$ is one of these $x_i$'s, and the associated $y_i$ is a double root of $V_{eff}(x_0,y)$ . The corresponding $z_i$ is determined by solving $f(x_i,y_i,z)=0$. Then one checks under which conditions each $x_i$ 
verifies $n(x_i,y_i,z_i)=2$ and finally take the greatest one.
Considering $x_0$, one has $y_0 = d_1, z_0 = 0$, and
$$
n(x_0,y_0,z_0)=1 + {d_1^2 \over d_6^2} + \sqrt{ {(d_1^2 - d_6^2)^2} \over d_6^4} =
\left\{
\begin{array}{ccc}
2 &\text{if}& d_1^2 \leq d_6^2,\\
2{d_1^2 \over d_6^2} > 2  & \text{if}& d_1^2 > d_6^2.
\end{array} \right.
$$
Therefore $x_0$ may be solution only if $d_1^2 \leq d_6^2$.
Likewise, with the corresponding $y_1$ and $z_1$ given in appendix, one checks that $x_1$ cannot be solution unless $C \leq 0$.
On the contrary, $x_2$ and $x_3$ may always be solutions for there are $y_2, z_2$ and $y_3, z_3$ such that 
$n(x_2,\,y_{2} ,\, z_2)=n(x_3,\,y_{3} ,\,z_3)=2$ under no particular condition.
By proposition 2, cancelling all links except $d_1$, $d(1,2) \leq x_0$. So if $d_1^2 \leq d_6^2$ then $d(1,2)=x_0.$ As $x_1 \geq x_2$ and
$x_1 \geq x_3$,  $d(1,2) = x_1$ if $C \leq 0$, otherwise $d(1,2)= \max(x_1,x_2)$.

When $d_1d_6=d_3d_4$, $x_1$ is not defined but the proof follws the same way.

Calculation  of $d(1,3)$ is the same, except we are searching the maximum of $y$. 
$Dis(V,x)$ is a polynomial in $y$ of degree twelve, with single roots $\pm y_0, \pm y_1$ and double roots $\pm y_2, \pm y_3$:
\begin{eqnarray*}
y_0 = {\sqrt{{{{d_3}}^2} + {{{d_6}}^2}}} &,& y_1 = {\sqrt{{{{d_1}}^2} + {{{d_4}}^2}}},\\
y_2 = {d_1}{\frac{{| {d_1}\,{d_3} +{d_4}\,{d_6}|}}{{\sqrt{{{( {d_3} + {d_4})}^2} +{{(d_1-{d_6}) }^2}}}}} &,&
y_3 = {\frac{{|d_1d_3+d_4d_6|}}{{\sqrt{{{( {d_3} - {d_4}) }^2} +  {{( {d_1} + {d_6}) }^2}}}}}.
\end{eqnarray*}
 With the associated $x_i,z_i$ given in appendix, one checks that $y_0$ 
(resp. $y_1$) may be solution if ${{{({{{d_3}}^2}+{{{d_6}}^2})}^2}
\leq{{( {d_3}{d_4}-{d_1}{d_6})}^2}}$ (resp. 
${{({{{d_1}}^2}+{{{d_4}}^2}) }^2}\leq{{({d_3}{d_4}- {d_1}{d_6})}^2})$.
As above, $y_2$ and $y_3$ may always be solution. Then, remark that $y_2, y_3 \leq y_0$ and $y_2, y_3 \leq y_1$. Finally, $y_0$ and $y_1$ cannot be simultaneous
distinct solutions, for adding both conditions yields $y_0=y_1$.$\blacksquare$ 
\bigskip

The four point space shows that there is no hope to find a general formula for the metric in any commutative finite spaces: distances cannot be read directly in the Dirac operator through a
finite algorithm. Computing the metric requires a more pragmatic approach and shall be undertaken case by case.

\section{Distances and axioms of noncommutative geometry.} 
In the previous discussion, we worked with triplets $(\aa,\hh,D)$ as if they satisfied all the axioms of noncommutative geometry. 
These axioms are introduced in order to recover the standard spin and riemannian geometries in 
the commutative case \cite{gravity}. Accordingly, for our distances to be bona fide noncommutative generalizations of riemaniann metrics, 
they have to be computed using triples satisfying all these axioms. 

However, these axioms lead, in the finite case \cite{krajew}, to matrices whose size increases rapidly with $n$ and thus prevents any 
computation except in few simple cases. This is the reason why we did not use these axioms up to now, but we shall see that the axioms do not put any constraints on the distances. 
$\\$

{\it {\bf Proposition 13.}
Let 
$\lp d_{ij}\rp_{1\leq i,j\leq n, i\neq j}$ 
be any finite sequence of possibly infinite strictly positive numbers such that 
$d_{ij}=d_{ji}$ and $d_{ij}\leq d_{ik}+d_{kj}$. Then there exists a spectral triple $(\aa,\hh,D)$ with $\aa=\mathbb{C}^{n}$ satisfying all 
the axioms, and such that the resulting distance on the set of pure states of $\aa$ is given by the numbers $d_{ij}$.}
$\\$

To proceed, we shall first prove the following lemma. 
$\\$

{\it {\bf Lemma 14.}
There is a spectral triple $(\aa,\hh,D)$ with $\aa=\mathbb{C}^{n}$ satisfying all the axioms such that
\begin{equation}
\label{comm}
\|[D,\pi(a)]\|=\mathop{\sup}\limits_{1\leq i,j\leq n, i\neq j}\frac{|a_{i}-a_{j}|}{d_{ij}}, 
\end{equation}
where $a=(a_{1},\dots,a_{n})\in\mathbb{C}^{n}$ and $\pi$ denotes the representation of $\mathcal{A}$ on $\mathcal{H}$.}
$\\$

{\it Proof.}
The proof is by induction on $n$.

The first non trivial case is $n=2$. We take $\aa_{2}=\mathbb{C}^{2}$ and $\hh_2=\mathbb{C}^{3}$. The representation $\pi_{2}$ and the chirality 
$\chi_{2}$ are both diagonal and are given by $\pi_{2}(x_{1},x_{2})=\mathrm{diag}(x_{1},x_{2},x_{2})$ and $\chi_{2}=\mathrm{diag}(1,-1,1).$
The Dirac operator $D_{2}$ and the charge conjugation $\jj_{2}$ are defined as 
$$
\dd_{2}=\left( 
\begin{array}{ccc}
0       &{1\over {d_{12}}}&0\\
{1\over {d_{12}}}&0       &{1\over {d_{12}}}\\
 0      &{1\over {d_{12}}}&0
\end{array} 
\right),\quad 
\jj_{2}=\left( 
\begin{array}{ccc}
0&0&1\\
0&1&0\\
1&0&0
\end{array} \right)C,
$$
where $C$ is the complex conjugation and we set ${1\over {d_{12}}}=0$ if $d_{12}=\infty$. 

In the finite case, all axioms reduce to the {\it reality}, {\it first order}, {\it orientability} and {\it Poincar\'e duality} axioms \cite{krajew}. 
In the present case, the first two are commutation relations easy to check due to the commutative nature of the algebra. 
The orientability axiom is fullfiled by writing the chirality as
$$\chi_{2}=\pi_{2}(1,-1)\jj_{2}\pi_{2}(-1,1)\jj_{2}^{-1}. $$

The multiplicity matrix is
$$
\mu_{2}=
\left( \begin{array}{cc}
0&1\\
1&-1
\end{array} \right)
$$
which is non degenerate and thus Poincar\'e duality holds. Finally. one easily checks that

$$
||\lb\dd_{2},\pi_{2}(x)\rb||=\frac{|x_{1}-x_{2}|}{d_{12}}. $$

Let us now assume that $(\aa_{n},\hh_{n},D_{n})$ together with $\pi_{n}$, $\chi_{n}$ and $\jj_{n}$ have been constructed for $n>2$. 
To build the order $n+1$ spectral triple, we merely imitate the $n=2$ construction. Let us take $\aa=\mathbb{C}^{n+1}$ and
$$
\hh_{n}=\hh_{n-1}\oplus \left( \mathop{\oplus}\limits_{i=1}^{n-1}\hh_{n}^{i}\right),$$
with $\hh_{n}^{i}=\mathbb{C}^{3},\, \forall i,n$. With respect to the previous decomposition all operators $\oo$ corresponding to $\dd,\pi,\chi$ and $\jj$ are block 
diagonal and defined inductively as
$$\oo_{n}=\oo_{n-1}\oplus\left(\mathop{\oplus}\limits_{i=1}^{n-1}\oo_{n}^{i}\right). $$
As in the $n=2$ case, we define
$$
\pi_{n}^{i}(x_{i},x_{n})=\mathrm{diag}(x_{i},x_{n},x_{n})\,, \; \chi_{n}^{i}=\mathrm{diag}(1,-1,1).
$$
The Dirac operator $D_{n}$ and the charge conjugation $\jj_{n}$ are defined as
$$
D_{n}^{i}=
\left( \begin{array}{ccc}
0       &{1 \over {d_{in}}}   &0\\
{1 \over {d_{in}}}&0          &{1 \over {d_{in}}}\\
 0      &{1 \over {d_{in}}}   &0
\end{array} \right),
\quad 
\jj_{n}^{i}=\left( 
\begin{array}{ccc}
0&0&1\\
0&1&0\\
1&0&0
\end{array} \right)C. $$
Then it is easy to check that all axioms but Poincar\'e duality hold using the induction assumption and the block diagonal nature of the construction. 

The multiplicity matrix of this spectral triple is 
$$
\mu_{n}=\left( 
\begin{array}{cccc}
0&1&\dots&1\\
1&-1&\ddots&\vdots\\
\vdots&\ddots&\ddots&1\\
1&\dots&1&-(n-1)
\end{array} \right)
$$
If $N$ is any positive integer, we can always consider the trivial spectral 
triple $(\mathbb{C}^{n},\mathbb{C}^{n}\oplus\mathbb{C}^{N},0)$ with obvious representation and charge 
conjugation and whose chirality is equal to -1. If we take the direct sum of 
this spectral triple with $(\mathcal{A}_{n},\mathcal{H}_{n},D_{n})$, the resulting 
multiplicity matrix is $\mu_n+NI_{n}$, which is non degenerate for $N$ 
sufficiently large. Accordingly, Poincar\'e duality will be satisfied.

Finally, the computation of the norm of the commutator $||\lb\dd_{n},\pi_{n}(a)\rb||$ follows easily from the block diagonal structure and 
the induction assumption. $\blacksquare$
$\\$

To complete the proof of proposition 13, we use the previous lemma to construct a spectral triple $(\aa,\hh,D)$ fulfilling condition 
(\ref{comm}). If $a$ verifies the norm condition, then $|a_{i}-a_{j}|\leq d_{ij}$, so $d(i,j)\leq d_{ij}$. 

Furthermore, if we fix any two points  
such that $d_{ij}<\infty$ 
and take $x_{k}=d_{ik}$ (which is finite thanks to triangular inequality) one has $|x_{i}-x_{j}|=d_{ij}$. An other application of the triangular 
inequality yields $|d_{ik}-d_{il}|\leq d_{kl}$ for any $k$ and $l$ so that $||\lb\dd,\pi(x)\rb||\leq 1$ by (\ref{comm}). This shows that 
$d_{ij}\leq d(i,j)$ so that the equality holds when $d_{ij}$ is finite. 

If $d_{ij}$ is infinite, so are $d_{ik}$ and $d_{jk}$ for any $k$. Thus, the inequality (\ref{comm}) does not constraint $x_{i}$ and $x_{j}$ since 
the corresponding matrix element of $D$ vanish.Therefore, we can send $|x_{i}-x_{j}|$ to infinity and we also have $d(i,j)=d_{ij}$. $\blacksquare$
$\\$

\section{ Conclusion.}

As a conclusion of previous discussion, we may say that once given $\mathcal{A}=\mathbb{C}^n$, there is no constraint arising from the 
axioms of noncommutative geometry. Such constraints may only appear if one imposes some extra conditions, such as fixing $\mathcal{H}=\mathbb{C}^n$ 
as we did in the discussion of the three and four point cases. We stress that we only showed that the map which associates a 
metric to a Dirac operator is surjective. In a discrete analogue of a quantum theory of gravity based on eigenvalues of the Dirac operators \cite{landi}, 
one also needs to know how many Dirac operators correspond to a given metric, as well as the possible relations between their spectra.

A naive question remains unanswered: what does these distances mean? According to spectral action principle
\cite{spectral}, Dirac operator encodes both physics and metrics. In the standard model, computation of the action leads to the lagrangien of the full
standard model and the Einstein-Hilbert action of general relativity. So the coding  of physics makes sense. But what about the coding  of metrics ?
For the time beeing, the answer is clear in the simple continuous case where noncommutative distance is the geodesic distance. Distance in the geometry of
standard model should have physical meaning, but it has not been explicitly computed yet. 
\pagebreak
\section*{Appendix.}
\subsection*{Coefficients of $V_{eff}(x,y)$ in the general four point case.}

\begin{eqnarray*}
V_4(x)&=& {\frac{4{{(d_3 d_4-d_2d_5)}^2} 
          ({{d_4}^2}d_6^2+ d_2^2 ({{d_4}^2}+ d_6^2))}{{ d_2^4}d_3^2d_4^4d_5^2d_6^2}},\\
V_3(x) &=&{\frac{8x(d_2d_5-d_3d_4)( d_3d_4d_5 d_6( d_2^2+ d_4^2) 
          + d_1(d_2d_3d_4(d_4^2+d_6^2) -d_4^2d_5d_6^2 -d_2^2d_5(d_4^2+2d_6^2)))}{ d_1{ d_2^3} d_3^2
          { d_4^4} d_5^2 d_6^2}},\\
V_2(x)&=& {{4{x^2}}\over{d_1^2d_2^2d_3^2d_4^4 d_5^2 d_6^2}}
[{d_4^2  (d_3^2d_4^2d_5^2+d_1^2{{(d_3d_4-d_2 d_5)}^2} 
+d_2^2( d_3^2d_4^2 + (d_3^2+d_4^2)d_6^2))}\\
      & &- 2d_1d_4d_6( d_2 d_4d_5(d_4^2-2d_3^2)+d_3d_4^2d_5^2
+d_2^2 d_3 (d_4^2 + 3d_5^2))\\
      & &+ d_6^2(d_4^2 {{(d_3 d_4-d_2d_5)}^2}+d_1^2(d_4^2(d_3^2+d_4^2+d_6^2)-6d_2d_3d_4d_5 +d_2^2(d_4^2+6d_6^2)))]\\
      &-& {\frac{4 (d_2^2(d_3^2 (d_4^2 +d_5^2 +d_6^2)+d_5^2 (d_4^2+2d_6^2))-2d_2d_3d_4d_5d_6^2  + d_4^2 (  d_5^2  d_6^2 +  d_3^2 (  d_5^2
+ 2  d_6^2 )  )  ) }{ d_2^2  d_3^2
          d_4^2 d_6^2 d_6^2}},\\ 
V_1(x) &=&{\frac{8{x^3}(d_1d_6-d_3d_4)(d_1d_2d_3d_4(d_4^2+d_6^2)-(d_1^2d_2d_4^2-d_3d_4(d_1^2+d_4^2)d_5+d_2(2d_1^2+d_4^2)d_6^2)d_6)}{{d_1^3}d_2d_3^2 { d_4^4}
d_6^2d_6^2}}\\
       &+&{\frac{8x(d_3d_4d_5d_6(d_3 d_4-d_2d_5) +d_1 ( d_2  d_3^2 (  d_4^2 +  d_6^2 )  + 
       d_6^2(d_2(d_3^2+2d_5^2)-d_3d_4d_5))) }{d_1 d_2  d_3^2  d_4^2  d_6^2  d_6^2}},\\ 
V_0(x)&=& 4 ({d_3^{-2}}+{d_6^{-2}}+{{d_6}^{-2}}) +{\frac{4{x^4}( d_4^2  d_5^2 +  d_1^2(d_4^2+d_5^2))  
      {{( d_3 d_4 - d_1  d_6 ) }^2}}{{ d_1^4}  d_3^2 { d_4^4}  d_6^2  d_6^2}} \\
      &-&{\frac{4{x^2}(d_4^2(2d_3^2d_5^2 +  d_6^2(d_3^2+d_5^2 )) -2d_1d_3d_4 d_5^2d_6  + 
         d_1^2( d_6^2(d_4^2 + 2d_5^2)+d_3^2 (d_4^2+d_5^2+d_6^2)))}{d_1^2d_3^2d_4^2d_6^2d_6^2}}.
\end{eqnarray*}
\subsection*{Computation of $d(1,2)$ when ${1\over d_2}={1 \over d_5}=\infty$.}
\begin{eqnarray*}
y_1 =\text{sign}(d_1d_6 - d_3d_4) d_6 \sqrt{{d_3^2 + d_6^2}\over {d_4^2 + d_6^2}} &,& 
z_1 =d_3{(d_1d_3 + d_4d_6)\over{\sqrt{(d_3d_4-d_1d_6)^2}}}\sqrt{ {d_4^2 + d_6^2}\over{d_3^2 + d_6^2}},\\
y_{2} = {\frac{d_1|{d_3}-{d_4}|\pm |d_4({d_1}+{d_6})|}{{\sqrt{{{( {d_3} - {d_4}) }^ 2} + {{( {d_1} + {d_6}) }^2}}}}}&,& 
z_{2} = \pm {\frac{{d_3}\,\left( {d_1} + {d_6} \right) }{{\sqrt{{{\left( {d_3} - {d_4} \right) }^2} +{{\left( {d_1} + {d_6} \right) }^2}}}}},\\
y_{3} = {\frac{d_1|{d_3}+{d_4}|\pm|d_4({d_1}-{d_6})|}{{\sqrt{{{( {d_3}+{d_4}) }^ 2} + {{( {d_1}-{d_6}) }^2}}}}}&,& 
z_{3} = \pm {\frac{{d_3}\,\left( {d_1}-{d_6} \right) }{{\sqrt{{{\left( {d_3}+{d_4} \right) }^2} +{{\left( {d_1}-{d_6} \right) }^2}}}}}.
\end{eqnarray*}
The choice of the signs depends on the sign of expressions in modules.
\subsection*{Computation of $d(1,3)$ when ${1\over d_2}={1 \over d_5}=\infty$.}
\begin{eqnarray*}
x_0 &=& {d_1d_6\sqrt{d_3^2 + d_6^2}\over{d_1d_6-d_3d_4}}, \; z_0 ={d_3^2\over\sqrt{d_3^2 +d_6^2}}, \\ 
x_1 &=& {d_1^2\over\sqrt{d_1^2 +d_4^2}}                 , \; z_1 ={{d_3d_4\sqrt{d_1^2 +d_4^2}}\over{d_3d_4 - d_1d_6}},\\
x_2 &=& \text{sign}(d_1d_3+d_4 d_6){\frac{d_1\,( d_3 + d_4 ) }{{\sqrt{{{( d_3 + d_4 ) }^2} + {{( d_1 - d_6 ) }^2}}}}},\\
z_{2}&=&{\frac{d_3(d_3{\sqrt{{{(d_1d_3+d_4d_6)}^2}}}\pm d_6(d_3(d_3+d_4)-d_1d_6+{{d_6}^2}))}
{{\sqrt{{{(d_3+d_4)}^2}+{{(d_1-d_6)}^2}}}({{d_3}^2}+{{d_6}^2})}},\\
x_3 &=& \text{sign}(d_1d_3+d_4d_6){\frac{{d_1}({d_3}-{d_4})}
{{\sqrt{{{({d_3}-{d_4})}^2}+{{({d_1}+{d_6})}^2}}}}},\\
z_{3} &=&{\frac{d_3(d_4{\sqrt{{{(d_1d_3+d_4d_6)}^2}}}\pm d_6(d_4(d_4-d_3)+d_1(d_1+d_6)))}{(d_3 d_4-d_1d_6){\sqrt{{{( d_3-d_4)}^2}+{{(d_1+d_6)}^2}}}}}.
\end{eqnarray*}
The choice of the signs depends on the sign of expressions in modules.

\pagebreak

\end{document}